# Unitary Representations of the Translational Group Acting as Local Diffeomorphisms of Space-Time

Moffat J*, Oniga T and Wang CHT

*Department of Physics, University of Aberdeen, King's College, Aberdeen AB24 3UE, UK*

**Abstract**

We continue to develop further a new mathematical approach to the quantisation of general field theories such as general relativity and modified gravity. Treating quantum fields as fibre bundles, we discuss operators acting on each fibre that generate a 'Fibre Algebra'. The algebras of two types of operators are considered in detail, namely observables as generic physical variables and more specialised quantum operators suitable for describing particles, symmetries and transformations. We then introduce quantum states of these operators and examine their properties. By establishing a link between the commutativity and group cohomology of the translational group as a local gauge group, we show that this leads to unitary representations. of the local gauge group of diffeomorphisms under very general topological conditions; as well the construction of generalised symmetric quantum states invariant under this group action. Discussion of these results in the context of loop quantum gravity and other current theories highlights constraints on the local nature of space-time.

## Introduction

The development of perturbation methods in quantum field theory by Dyson, Feynman and others, was an essentially *ad hoc* answer to the problems generated by the process of taking renormalisation limits for point particles. The further development of the renormalisation group based on ideas of self-similarity, has led us to develop new approaches to the quantisation of paths in spacetime [1,2].

One way of eliminating the infinities of renormalisation is through assuming a form of supersymmetry (SUSY) in which additional 'spinor charge observables' are added to the Lie algebra of the Poincaré group of spacetime symmetries. In an irreducible representation of the resulting 'graded algebra', the centre of the algebra is reduced to multiples of the identity operator. It turns out that all particles in this representation have equal mass, as in the Wigner theory of irreducible representations of the Poincaré group and its Lie algebra [3]. However, the spins of the particles in this 'super'-representation are not fixed at a common value. In other words, the representation corresponds to a multiplet of particles, all with the same mass, but with different spins (equal numbers of fermions and bosons called partners and superpartners). At a given perturbation loop level, this generates, in a naïve application of the theory, equal mass contributions to the calculation but with loop contributions of opposite sign which cancel, giving a finite result. In practice, this symmetry is broken, resulting in different masses for the partners and superpartners, giving only approximate cancellation [4]. This relates to non-commutative spacetime and gravity through the superfield formalism [5-7].

Theories successfully describing the electroweak and strong forces take a similar approach. They use a group action acting on the Lagrangian which is symmetric. However, in contrast to global SUSY, this gauge group action is assumed to be local. If we apply this approach to SUSY, in the linear representation, the translation operator is represented by $P_\mu = \partial_\mu = \partial/\partial x^\mu$ corresponding to local translations in space-time between arbitrary coordinate frames i.e., a theory of 'supergravity' (SUGRA). We expected that this gravitational theory should involve a spin-2 graviton and its fermionic spin 3/2 gravitino companion particle. However, initial survey results at CERN using the TLAS detector [8], confirmed by recent postings, indicate no significant deviation from the Standard Model at 13 TeV, implying that the mass symmetry is badly broken; the mass-energies of the superpartners (if they exist) are possibly much larger than initially thought and limit their ability to control loop divergence. This leaves open the possibility of other potentially finite but more algebraic approaches including noncommutative geometry [9,10], and modified theories of gravity such as scalar-tensor gravitation [11-13].

The principle of local gauge symmetries inherent in interacting quantum fields is indeed a fundamental geometric origin of nonlinearities that are difficult to treat with a perturbation theory. Since the starting point of such a theory normally requires a background configuration with gauge fixing that could not only limit but also potentially affect the consistency of quantum dynamics. In contrast, gauge principle offers a natural underlying conceptual simplicity especially formulated in terms of modern fibre bundles, at the expense of introducing symmetry-redundant degrees of freedom.

The canonical quantisation of interacting systems without artificially restricting gauge symmetries have been pioneered by Dirac [14,15] in terms of an extended Hamiltonian formalism with 'first class' constraints that distinguish dynamical degrees of freedom while preserving and generating gauge transformations. When the extended Hamiltonian analysis is applied to general relativity, the Hamiltonian of gravity turns out to be completely constrained in terms of the Hamiltonian constraint and the momentum constraints [16]. The latter are also known as the diffeomorphism constraints since at least classically they generate spatial diffeomorphisms. There are two possible ways to 'process' these constraints. The first procedure uses them to generate the quantum dynamics of the spatial geometry using Dirac quantisation [17,18]. The second procedure is based on the reduced phase space method [19-21], where all constraints would be eliminated at the classical level in order to obtain a nonvanishing effective Hamiltonian that generates the evolution of certain unconstrained geometric variables. As discussed above, in this process, gauge is fixed classically and so it is not clear whether any resulting quantisation would preserve the full gauge properties of general relativity.

The conceptual and technical benefits of Dirac's constraint quantisation have recently been demonstrated through applications in gauge invariant interactions between generic matter systems and their weakly fluctuating gravitational environments leading to quantum decoherence of matter and spontaneous emissions of gravitons [22-27]. There have also been ongoing efforts to achieve full Dirac quantisation of general relativity in a background independent and non-perturbative approach using loop quantum gravity [28-31,17].

Our aim, following Dirac, is to initiate and develop a finite mathematical framework based on a new algebraic approach to quantum field theory and quantum gravity derived from foundational work by Moffat [32-36] and later developments including [1,2]. We specifically address the question of diffeomorphism invariant quantum states as the diffeomorphism constraints in the current formulation of loop quantum gravity could not be implemented through unitary representations. While this work does not invoke a particular approach to quantum gravity, such as loop quantisation, we discuss possible future prospects of incorporating our present development in the context of current quantum gravity research.

## Algebras of Quantum Observables

A commutative operator algebra of observables is equivalent to the set of continuous functions on a compact Hausdorff space. This equivalence arises through the Gelfand transform of a quantum operator $A$, an observable in a set of commuting observables. This maps $A \to \hat{A}$ with $\hat{A}(\rho) \to \rho(A)$ and $\rho$ a continuous complex valued homomorphism.

Since the Gelfand transform $A \to \hat{A}$ is an algebra isomorphism, the spectrum (the set of measurement eigenvalues) of the operator $A$, denoted $(A)$, equals $\sigma(\hat{A})$. More generally we take as general context for this the modern algebraic theory of relativistic quantum fields [1,2,37,38] in which we denote the set of all 'local observables' $O(D)$ as representing physical operations performable within the space-time constraints $D$. In particular, if $O(D)$ is weakly closed (i.e., a von Neumann algebra) and $A$ is an observable in $O(D)$ then we assume that $A$ is both bounded and self-adjoint, thus $A$ generates an abelian subalgebra of $O(D)$ containing its spectral projections corresponding to the measurement process.

If $g \to \alpha_g$ is the corresponding mapping from the Poincare group $P$ to the group of all automorphisms of $O(D)$ then $\alpha_g(O(D))=O(gD)$. We also assume that $D_1 \subseteq D_2$ implies $O(D_1) \subseteq O(D_2)$. The union of all the local algebras generates an algebra of all observables. The closure in the ultraweak operator topology generates the 'quasi-local' von Neumann algebra of all observables.

The idea behind [9,39] consists of finding non-commutative generalisations of the structure of the Einstein-Hilbert action which can be applied in order to gain greater insight into the Standard Model action. The approach taken introduces the theory of fibre bundles. A *fibre bundle* is a pair $(E,\pi)$ where $E$ is a topological space and $\pi$ is a projection onto a subset $M$ called the base space. For a point $x$ in $M$, the set $F=\{y \in E \cap \pi^{-1}(x)\}$ is the fibre at the point $x$, and it is assumed that the bundle is locally a product space about each point $x$. $E$ is a vector bundle if each fibre has the structure of a vector space. In this case we can go further and define the fibre at $x$ as consisting of all bases of the vector space attached to the point $x$; and define the group $G$ of changes of basis, acting on the fibre space $F$. This is an example of a 'principal bundle'. A section or lifting $s$ through the fibre bundle $E$ with projection function $\pi$ mapping $E$ to the base space $M$ is a continuous right inverse for $\pi$.

We thus assume that each local algebra $O(D)$ should correspond to the algebra of sections of a principal fibre bundle with base space a finite and bounded subset of spacetime, such as the interior of a bounded double-cone [38]. The algebraic operations of addition and multiplication are assumed defined fibrewise for the algebra of sections. Additionally, the group of unitaries $U(x)$ acting on the fibre Hilbert space is our gauge group. For example if the fibre $F(x)$ has a two-component basis, as a Hilbert space, of Clifford variables; then the spinor structure of left and right handed Weyl 2-spinors, and a gauge group including SU(2), form the basis of our current understanding of electroweak unification.

At the algebra level, this representation consists of the section algebra $\mathbf{A}(x)$ of 2×2 matrix operators acting on the fibre spinor space $F(x)=\pi^{-1}(x)$ with fibrewise multiplication and addition. It includes all such matrices and is thus a von Neumann algebra which is both norm separable (equivalently, finite dimensional) and with a trivial centre. We assume that this applies more generally and define such a norm separable matrix algebra with trivial centre to be a *quantum operator algebra*.

Let us now introduce a *quantum state of* $\mathbf{A}$ to be a norm continuous linear functional $f$ acting on $\mathbf{A}$ which is real and positive on positive operators and satisfies $f(I)=1$. To distinguish these from the more restricted vector states of the form $\omega_x(A)= \langle x, Ax \rangle$, in Dirac notation, we may refer to $f$ as a *generalised* quantum state. The state space is a convex subset of the Banach dual space $\mathbf{A}^*$ of $\mathbf{A}$ and thus inherits the weak * topology from $\mathbf{A}^*$. The unit ball of $\mathbf{A}^*$ is weak * compact and the state space is a closed and thus compact, convex subset of the unit ball. Application of Krein and Milman [40] shows that it is the weak * closed convex hull of its extreme points. These extreme points are called pure quantum states. A normal state is a state which is also continuous for the ultraweak operator topology. The ultraweak topology on the algebra $\mathbf{A}$ can in fact be derived from the h of $\mathbf{A}$ as the dual space of the set of normal states of $\mathbf{A}$, so that, formally, $(A,f)=(f,A)$ for each element $A$ of $\mathbf{A}$ and each normal state $f$; i.e., $A$ is a non-commutative Gelfand transform [41-44].

## The Translational Group Action on the Fibre Algebra

The fibre algebra $\mathbf{A}(x) = \{A(x); A(x)$ is an operator acting in the Hilbert space $F(x)\}$ is assumed to be a quantum operator algebra, and we select as our local gauge group the translational group $T$ acting as a group of automorphisms of $\mathbf{A}(x)$. The translational group $T$ is a subgroup of the Poincaré group $P$, a locally compact, connected Lie group, and is generated by the 4 basic linear translations in space-time plus 6 rotations and boosts. These define a smooth manifold, and each element $g$ of $P$ is of the Lie group form:

$$g = T(\varphi_1, \cdots, \varphi_n)$$

where $T$ is the corresponding point on the manifold surface, and with $n=10$ parameters in our case. The origin of the manifold $T(0,0,¼,0)$ corresponds to $e$, the group identity, denoted $id$.

Denote $\phi=(\phi_1,,\phi_n)$ and assume $\phi$ is a first order infinitesimal. This means that $\phi^2$ is negligible, and that $T(\phi)$ is close to the origin $e$ of the group. If $U$ is a unitary representation of $P$ on a Hibert space H then to first order:

$$U(g) = U(T(\varphi)) = I + i\pi(\varphi^\mu t_\mu) = I + i\varphi^\mu \pi(t_\mu)$$

with $\pi$ a mapping from the Lie algebra of generators $t_\mu$ of $P$ into the set of linear operators acting on $H$. $U(g)$ a unitary operator implies that $\pi(t_\mu)$ is a Hermitian linear operator, an observable.

Additionally, let $\phi$ become a second order infinitesimal so that $\phi^2$ is not negligible but $\phi^3$ is negligible. Since $U$ is a group representation we have $U(g_1)U(g_2) = U(g_1 g_2)$. From this expression, it follows that:

$$\pi[t_\alpha, t_\beta] = i f^\gamma_{\alpha\beta} \pi(t_\gamma)$$

with $f$ the antisymmetric structure constants of the group [45-48]. This then gives rise to a representation $\pi$ of the corresponding Lie algebra generators. Consider now the translational group $T$ acting locally as unitaries $U(g)$. From ref. [38] and our previous discussion, we know that if $g$ is an element of the translational group $T$ then $g$ has the form $g = \exp(-iP(g)\cdot a)$ where $a$ is the corresponding group parameter, $P(g)$ is the Lie algebra generator of $g$, and $P(g)\cdot a = P_\mu(g)a^\mu$. Hence under the homomorphism $\pi$, $U(g) = \exp(-i\pi(P(g)\cdot a))$.

## Group Extensions and Mackey Theory

A group $G$ is called an extension of a group $C$ by a group $B$ if $C$ is a normal subgroup of $G$ and the quotient group $G/C$ is isomorphic to $B$. We assume in what follows that the group $C$ is abelian. These assumptions imply [49] that the sequence:

$$0 \to C \to G \to B \to 0$$

is a short exact sequence with the mapping from $C$ to $G$ being the identity map, and from $G$ to $B$ being (up to isomorphism) the quotient mapping from $G$ to $G/C$. The converse is also true in the sense that any short exact sequence of groups can be identified with the extension of a group.

Given an abelian group $K$ (say), and an arbitrary group $Q$, we assume that we have mapping $\gamma$ and $\eta$ such that : $Q \to \text{aut}(K)$ is a group homomorphism from $Q$ to the group of all automorphisms of $K$ and $\eta$: $Q \times Q \to K$ is a system of factors for $Q$ and $K$. Together these two mappings can be used to define a multiplication on the set $K \times Q$ giving it a group structure:

$$(\xi_1, y_1)(\xi_2, y_2) = (\xi_1 \gamma(y_1)(\xi_2) \eta(y_1, y_2), y_1 y_2)$$

$\forall \xi_1, \xi_2 \in K, y_1, y_2 \in Q$. Defining $\theta: \xi \to (\xi, e)$ and $\theta: \xi \to (\xi, e)$ with $e$ the identity of $Q$ then $K \times Q$ with this group structure is an extension of $K$ by $Q$, normally denoted as $K\eta Q$. It follows that the sequence:

$$0 \to K \xrightarrow{\theta} K \times Q \xrightarrow{\delta} Q \to 0$$

is a short exact sequence. If $Q$ and $K$ are groups with borel measure structures, and $\eta$ is borel mapping then we say that $K\eta Q$ is a borel system of factors for $Q$ and $K$. Given these conditions we have the following result [50].

There exists in the group extension $K\eta Q$ a unique locally compact topological structure relative to which $K\eta Q$ is a topological group such that the identity map from $K \times Q$ (as a product borel space) into $K\eta Q$ is a borel measurable mapping. The map $\theta$ is a bicontinuous isomorphism from $K$ onto a closed normal subgroup of $K\eta Q$. The isomorphism from $K\eta Q/\theta(K)$ onto $Q$ defined by the mapping $\delta$ is also bicontinuous. Moreover, $K\eta Q$ is a separable group.

We will refer to the above as Mackey's theorem. An early version of this result was conjectured by Moffat [34] in relation to representations of the Poincare group.

If $\alpha:g \to \alpha_g$ is a group representation of the Poincaré group $P$ as automorphisms of the quantum operator algebra $\mathbf{A}$ and there is for each $g \in P$ a unitary $U_g$ implementing $\alpha_g$ then $U_g U_h$ and $U_{gh}$ both implement $\alpha_{gh} = \alpha_g \alpha_h$. There is thus a phase factor $\lambda(g,h)$, a complex number of modulus 1, with $U_g U_h = \lambda(g,h) U_{gh}$. By considering in a similar way the group product $ghj=(gh)j=g(hj)$ these phase functions $\lambda(g,h)$ satisfy the requirement to be 2-cocycles of $P$.

No that if we set $V_g = v(g)U_g$ with $|v(g)|=1$ then for $g \in P$, $V_g$ also implements $\alpha_g$. We define a lifting as the choice of an appropriate $v(g)$ so that $g \to v(g)U_g$ is a group representation i.e., the 2-cocyles are trivial. In cohomology terms the expression $v(g)v(h)/v(gh)$ defines a coboundary. A local lifting is a lifting which is restricted to a neighborhood of the identity of the group.

The Poincaré group $P$ is locally Euclidean and the underlying topology is locally compact. The Poincaré group was the first locally compact non-abelian group in which the later Mackey methodology was initially developed by Wigner in his theory of unitary irreducible representations of $P$. [3,44]. The group thus has an associated Haar measure $m$. The translation subgroup $T$ and each of the rotation groups $R(x^\mu)$ around a single spacetime axis $x^\mu$ are abelian path-connected subgroups of $P$.

A representation of group $T$ as automorphisms of a quantum operator algebra $\mathbf{A}(x)$ is a group homomorphism $\alpha:g \to \alpha_g$ from $T$ into aut$(\mathbf{A}(x))$, the group of all automorphisms of $\mathbf{A}(x)$. This representation is *norm measurable* (resp. *norm continuous*) if the real valued mapping $g \to \|\alpha_g - i\|$ is Haar-measurable (rep. continuous) as a function defined on $T$. From the Appendix, a weakly measurable representation of this form is norm measurable.

We have the following result:

Let $g \to \alpha_g$ be a group representation of the translational group $T$ as automorphisms of the quantum operator algebra $\mathbf{A}(x)$ which is norm measurable. Then the mapping $g \to \|\alpha_g - i\|$ is norm continuous i.e., the representation is norm continuous.

*Proof.* From the Appendix we can infer that the set $W$ defined below is measurable, since the representation is norm measurable:

$$W = \left\{ g \in T, \|\alpha_g - i\| \leq \frac{\varepsilon}{2} \right\}.$$

Each * automorphism $\alpha_g$ is an isometric linear map, thus if $g \in W$ then from the illustrative calculations in the Appendix, with $\alpha(g)=\alpha_g$ we have

$$\|\alpha_{g^{-1}} - i\| \leq \|\alpha_g - i\|.$$

Hence, if $g \in W$ then $g^{-1} \in W$, and if $g, h \in W$ then, exploiting again the isometry of each automorphism, from the Appendix we have:

$$\|\alpha_{gh} - i\| \leq \|\alpha_h - i\| + \|\alpha_{g^{-1}} - i\| \leq \varepsilon.$$

Thus we have

$$W^2 \subset \left\{ g \in T; \|\alpha_g - i\| \leq \varepsilon \right\}.$$

Now $W$ contains the identity $id$ in the Lie group $T$ and if $g \in W$, $g$ not equal to $id$, then $g=\exp(\theta k)$ where $K$ is an element of the Lie

Algebra. Thus $W$ contains intervals of the form $g(\theta)=\{\exp(\theta k), 0<\theta \leq \theta_1\}$ and, it follows, must have strictly positive measure. Since $T$ is locally compact, there is a compact set $C \subset W$. Then the set $CC^{-1}$ contains a neighbourhood $N$ of $id$. [45]

It follows that $N \subset CC^{-1} \subset W^2 \subset \{g \in T; \|\alpha_g - i\| \leq \varepsilon\}$ and the mapping $g \rightarrow \|\alpha_g - i\|$ is continuous at the origin. By translation it is continuous everywhere on $T$.

Although not assumed in the above result, typically, these automorphisms $\alpha_g; g \in T$ will be inner, and they will be generated by a unitary (gauge) representation $g \rightarrow W_g$ in the fibre algebra $A(x)$. In this case, since $A(x)$ is norm separable, it follows that the subset $S = \{W_g; g \in T\}$ is also norm separable, and we can select a countable dense subset $\{W_{gn}; n \in Z\}$ of $S$. Defining $UW$ as the neighbourhood

$$UW = \left\{ W_g; g \in T, \|W_g - I\| \leq \frac{\varepsilon}{2} \right\}$$

of the unitary gauge group in $A(x)$ complementary to the neighbourhood $W$ of the automorphism group, we can choose, for each group element g in $T$, a unitary $W_{gn}$ with $\|W_g - W_{gn}\| \leq \varepsilon/2$. This implies that

$$\|W_{gn^{-1}g} - I\| = \|W_g - W_{gn}\| \leq \frac{\varepsilon}{2}$$

and thus $(gn)^{-1} g \in UW$ and hence $T = \bigcup_n (gn)(UW)$. From this, it follows that the neighbourhood $UW$ also has positive measure.

A *unitary representation* the translational group $T$ is defined to be a group homomorphism $U: g \rightarrow U_g$ from $T$ into the group of unitary operators acting on the fibre Hilbert space $F(x)$. Let $\alpha$ be a group representation $g \rightarrow \alpha_g$ of $T$ as automorphisms of the quantum operator algebra $A(x)$ acting on $F(x)$. Then a unitary representation $U: g \rightarrow U_g$ of $T$ implements $\alpha$ if $\alpha_g(A) = U_g^* A U_g$ for all g in $T$ and all $A$ in $A(x)$.

We have the following key result:

Let $\alpha: g \rightarrow \alpha_g$ be a representation of the translation group $T$ as automorphisms of the quantum operator algebra $A(x)$, which is norm measurable. Then $\alpha$ is implemented by a norm continuous unitary representation $U: g \rightarrow U_g$ from $T$ to the group of unitary operators acting on the Hilbert space $F(x)$.

*Proof.* If $\alpha: g \rightarrow \alpha_g$ is a group representation of the translation group $T$ as automorphisms of the quantum operator algebra $A(x)$, which is norm measurable, then we already know that $\alpha: g \rightarrow \alpha_g$ is norm continuous. The translation subgroup $T$ is thus is an abelian path-connected, hence connected, group with a norm continuous representation as automorphisms of the von Neumann algebra $A(x)$. Thus by the main theorem in Moffat the result follows [35].

This is an extremely powerful result due to its generality. To briefly discuss:

- The original result [34] is valid in fact for any norm continuous automorphic representation of a connected abelian topological group acting on a general von Neumann algebra. There is aways a norm continuous unitary implementation in this case.

- The specific application to a weakly measurable (hence strongly continuous representation of the translational group $T$, comes about because of the result that any such automorhic representation is actually norm continuous. Topology goes hand in hand with algebra.

- This generality allows its application to (for example) the weak closure of any GNS representation derived from a generalised quantum state.

- The key step in the original proof was the novel application of ultrafilter theory to generate a lifting from the carrier space of the centre to the carrier space of a set of comuting unitaries. this exploits the duality between an observable and its Gelfand transform.

From the our discussion so far, we can also deduce the following;

Suppose there is a homomorphism $\pi$ from the Lie algebra of $T$ into $A(x)$ and a corresponding representation $\tau$ of the translational group $T$ as automorphisms of the quantum operator algebra $A(x)$. Then the representation $\tau: g \rightarrow \tau_g$ can be implemented by a strongly continuous unitary representation.

*Proof.* A strongly continuous automorphic representation of the translational group $T$ is weakly measurable. Thus there is a norm continuous unitary representation implementing $\tau$. A norm continuous unitary representation is certainly strongly continuous, and the result follows.

Examination of the proof developed by one of us (Moffat) in ref. [35] on which this and the previous result depend, indicate in fact a two stage process at play.

Step 1. Exploit the spectral properties of a set of implementing unitaries to show that these unitaries commute, generating a commutative algebra.

Step 2. Use the Gelfand transform to select out a 2-cocycle set giving a unitary representation.

## Discussion of Group Extensions and the Local Flatness of Spacetime

In our second result above, $W_g$ has the form $W_g = \exp(-i\pi(P(g).a))$. It follows that $\{W_g; \forall g \in T\}$ is a commuting set if and only if the translational group $T$ is abelian (i.e., spacetime is locally flat).

Let $N = Ker(\tau) = \{g \in T; \tau_g = id$ the identity$\}$. Replacing $T$ by $T/N$ we may assume that $\tau_g = \tau_h$ implies $g = h$.

Let $O$ be the unit circle. Since $A(x)$ is a quantum operator algebra, the unitary group of the centre of $A(x)$ is isomorphic to $O$. It is then easy to see that $\Gamma = \{\lambda W_g; \lambda$ is in the unit circle, and $g \in T\}$ is an abelian subgroup of the unitary group of $A(x)$.

Define the mapping $\gamma(\lambda) = \lambda I$ which maps the unit circle $O$ into the subgroup $\Gamma$, and the mapping $\eta(\lambda W_g) = g$ which maps the group $\Gamma$ to the translational group $T$. Then the sequence:

$$0 \rightarrow O \xrightarrow{\gamma} \Gamma \xrightarrow{\eta} T \rightarrow 0$$

is short exact. . Thus $\Gamma$ is an extension of $O$ by $T$. We can in fact identify the group $\Gamma$ with [45]

$\eta'(\lambda, g) = \eta(\lambda W_g) = g$

Since the Hilbert space $F(x)$ on which the unitaries act is separable, by assumption: it is sufficient to demonstrate that the mapping $g \rightarrow \langle \xi, U_g \xi \rangle$ is a borel measurable mapping for all $g \in T, \xi \in F(x)$. From the discussion above we can identify the group $\Gamma$ with the borel system of factors $O\eta'T$.

Let the mapping $J$ denote the identification $O \times T \leftrightarrow \Gamma \leftrightarrow O\eta'T$ then $J$ is a borel mapping by Mackey's theorem, from our earlier discussion of group cohomology. Thus the mapping $W: g \rightarrow W_g$ is a borel measurable mapping from the translational group $T$ into the group $\Gamma$. Hence the mapping $g \rightarrow \langle \xi, W_g \xi \rangle$ is a measurable mapping for any $\xi$ in the Hilbert space $F(x)$. In addition, the character $\sigma: \Gamma \to C$ is a continuous mapping, thus $\sigma \circ W : g \to \sigma(W_g)$ is borel. Combining these together it follows that the mapping $g \to \langle \xi, U_g \xi \rangle = \sigma(W_g)^* \langle \xi, W_g \xi \rangle$ is borel measurable.

## Invariant States of the Algebra of Observables

In this section we change the focus to recall the overall structure of the von Neumann algebra $R$ of all observables, consisting of the weak closure of the sets of all finite unions of local algebras $O(D)$. The structure of $R$ is much more general than that of a quantum operator algebra. For example, $R$ may well have a non-trivial centre. We further investigate the structure of $R$ by examining its invariant states. Each section algebra consists of operators $\{A \in O(D); A|_{F(x)} = A(x) \text{ for all } x \in M\}$ which are quantum fields, and each such section algebra is a subalgebra of $R$.

Let $f$ be a generalised quantum state of the algebra $R$, and $\alpha: g \to \alpha_g$ be a group representation of the Poincaré group $P$ as automorphisms of $R$. The state $f$ is said to be Poincaré invariant if $f(\alpha_g(A)) = f(A)$ for all $g \in P$ and $A$ in $R$.

We call a *T-symmetry state*, a quantum state of $R$ which is invariant under the translational group $T$. In ref. [2] we characterised the condition for such states to exist. As discussed earlier, there is a vector of Lie algebra generators of these translations $a^\mu$ denoted by the abstract vector $P_\mu$. Acting on $R$, a translation $a^\mu$ maps the local algebra $O(D)$ into the local algebra $O(D+a^\mu)$. It is worth noting here that the vacuum state of a quantum system (the state of lowest energy) on the Minkowski flat background is in general translation invariant with constant $a^\mu$ [38]. Since translational group $T$ is the local gauge group for diffeomorphisms, in this sense we identify $T$-symmetry states as diffeomorphism invariant states. Such symmetry states turn out to be not uncommon, as we now prove.

This takes us to the final result:

Let $\tau: g \to \tau_g$ be a group representation of translational group $T$ as automorphisms of the algebra of observables $R$ which is weakly measurable. Then there is always at least one quantum state of the algebra which is a $T$-symmetry state.

*Proof.* Let $f$ be a state of $R$ and define the 'induced' transformation $v_g(f) = f \circ \tau_g$. This means that $v_g(f)$ is also a state and $v_g(f)(A) = f(\tau_g(A))$.

Since the subgroup $T$ is abelian, and the mapping $g \to v(g)f$ is a group homomorphism, the set $\{v(g); g \in T\}$ is a continuous group of commuting transformations of the dual space $R^*$. If $f$ is a quantum state of the algebra then define $E$ to be the weak $^*$ closed convex hull of the set $\{v(g)f; g \in T\}$. Then $E$ is a weak $^*$ compact convex set and each $v(g): E \to E$. By the Markov-Kakutani fixed point theorem it follows that $E$ has an invariant element. Thus each quantum state generates a symmetry state which is in the closed convex hull of the set of all translations of $f$ [46].

## Concluding Remarks

Motivated by modern diverse developments of quantum field theory, quantum gravity, and modified gravity, including supersymmetry, supergravity, noncommutative geometry, scalar-tensor gravity, and loop quantum gravity, we have initiated a new algebraic approach to background independent quantisation of gravity and matter as a potential unified approach to quantising gauge theories. Here we specifically address the existence and construction of diffeomorphism invariant quantum states. For this purpose, an analysis of the group cohomology of the translational group as the local gauge for diffeomorphisms is carried out that proves to provide useful tools for the presented analysis.

Crucially, by establishing a link between the commutativity and group cohomology of the translational group, we show that a strongly continuous representation of the translational group can be implemented by a norm, hence strongly continuous unitary representation. This result leads to the construction of quantum states invariant under the action of the translational group, as the local gauge group of diffeomorphisms, with unitary representations. Applied to the GNS representation of an invariant state, many of which exist, as we have shown, this result leads to the construction of quantum states invariant under the action of the translational group, as the local gauge group of diffeomorphisms, with strongly continuous unitary representations.

It will be of interest to apply our framework to specific quantum gravity programmes involving further gauge properties, such as spin symmetry in standard loop quantum gravity and additional conformal symmetry, which may involve additional scalar fields, in conformal loop quantum gravity [17,47-49,31]. The details of such extended investigations and resulting invariant states relevant for the non-perturbative quantisation of general relativity, modified gravity, and gauge theories [28,30] will be reported elsewhere [50-56].

## Appendix

### Derivations related to weakly measurable representations

Assume that $A(x)$ is a norm separable quantum operator algebra acting on a separable fibre Hilbert space $F(x)$ and $T$ is the translational group, acting as automorphisms of $A(x)$.

Define $\xi(g): g \to \|\alpha_g - i\|$. $A(x)$ is finite dimensional, corresponding to an energy cut-off, and is norm separable, thus $\xi(g)$ is the pointwise limit of mappings of the form

$$\{\xi(g)(T_n): g \to \|\alpha_g(T_n) - T_n\|\}$$

Where $\{T_n\}$ is a countable dense subset of $A(x)$.

For each $n$ we have
$$\|\alpha_g(T_n) - T_n\|^2 = \sup_{\|x\|^2 \leq 1} \{\omega_x(\alpha_g(T_n) - T_n)^*(\alpha_g(T_n) - T_n)\}$$

is the pointwise limit of mappings of the form $\omega_{x(m)}(\alpha_g(T_n) - T_n)^*(\alpha_g(T_n) - T_n)$, since the Hilbert space is separable. If this mapping is measurable, it follows that the mapping $g \to \|\alpha_g(T_n) - T_n\|$ is Borel measureable and a weakly measurable representation of this type is norm measurable.

### Calculations related to the measurable set $W$

$$\|\alpha(h^{-1}) - i\| = \|\alpha(h^{-1})i - \alpha(h^{-1})\alpha(h)\|$$

$$\leq \|i - \alpha(h)\| = \|\alpha(h) - i\|$$

$$\|\alpha(gh) - i\| = \|\alpha(g)\alpha(h) - \alpha(g)\alpha(h)\alpha(h^{-1})\alpha(g^{-1})\|$$

$$\leq \|\alpha(h) - \alpha(h)\alpha(h^{-1})\alpha(g^{-1})\|$$

$$= \|\alpha(h) - \alpha(hh^{-1}g^{-1})\|$$

$$= \|\alpha(h) - \alpha(g^{-1})\|$$

$$\leq \|\alpha(h) - i\| + \|i - \alpha(g^{-1})\| \leq \varepsilon$$

**Acknoledgment**

We are grateful for financial support to the Carnegie Trust for the Universities